\documentclass[letterpaper, 10 pt, journal, twoside]{IEEEtran}
\usepackage{amsmath,amsfonts,amssymb,enumerate}
\usepackage{remark}
\newremark{remark}{Remark}

\newtheorem{lemma}{Lemma}
\newtheorem{theorem}{Theorem}

\newtheorem{corollary}{Corollary}
\newtheorem{example}{Example}
\usepackage{mathptmx} 

\usepackage{boxedminipage}
\usepackage{graphicx}
\usepackage{pst-sigsys}
\usepackage{pst-node,pst-circ}
\definecolor{dgray}{gray}{0.6}
\definecolor{lgray}{gray}{0.8}
\newcommand{\ignore}[1]{}

\newcommand{\F}{F}
\newcommand{\tF}{\tilde{F}}
\newcommand{\G}{G}

\newcommand{\tGone}{\tilde{G}_{1}}
\newcommand{\hA}{\hat{A}}
\newcommand{\tA}{\tilde{A}}
\newcommand{\tBone}{\tilde{B}_{1}}

\begin{document}


\title{\textbf{A Note on Order and Index Reduction for Descriptor Systems}}
\author{Martin J. Corless and Robert N. Shorten
\thanks{ M. J. Corless is  with the School of Aeronautics and Astronautics, Purdue University, West Lafayette, IN 47906 USA (e-mail:   corless@purdue.edu). 
R. N. Shorten is with the Dyson School of Engineering Design, Imperial College London, UK (email: r.shorten@imperial.ac.uk).
Corresponding Author: M. J. Corless}
}

\maketitle

\begin{abstract}
We present order reduction results for linear time invariant descriptor systems.
Results are given for both forced and unforced systems as well methods for constructing 
the reduced order systems. 
Our results establish a precise connection between classical and 
new results on this topic, and lead to an elementary construction of  quasi-Weierstrass forms for a descriptor system.
Examples are given to illustrate the usefulness of our results. 
 \end{abstract}
 
\thispagestyle{empty}
	\pagestyle{empty}

\parindent0mm

\section {Introduction}

Descriptor systems have been widely studied in the mathematics and engineering literature for several decades 
 \cite{CampbellBook1980,KunkelMehrmann2001Book,trenn2009}. Recently, they have also become very popular in
 the mainstream control engineering literature, especially in the context of switching and hybrid dynamical systems 
\cite{liberzon2009stability,zhai2011commutation, zhou2013, wirth2015, {strenn2012}, {trenn2012}, {Benner2019}}, motivated
in part, by the fact that descriptor systems provide a natural framework to model and 
analyse many dynamic systems with algebraic constraints (for example, a mechanical system with coordinate constraints) \cite{Sajjaal2019}. Formally, a  {\em descriptor} characterization of a dynamical system consists of a combination of differential equations and algebraic equations,
that coupled together describe the dynamics of the system under study.  Even though this formalisation is convenient
for many physical and man-made dynamic systems, the analysis of such systems requires bespoke 
techniques when compared with conventional systems. Our interest in this paper concerns linear time invariant descriptor systems,
and methods for characterising the qualitative properties of these systems in terms of lower order systems. As a special case we  
also consider reduction methods that yield a standard system; that is, a 
system described only by standard differential equations and no algebraic equations. Our motivation is deriving these tools is that
reduced order characterisations are often useful than the corresponding original descriptor characterisations due to their compatibility with the 
broad portfolio of existing results in Systems Theory which characterise the properties of ordinary differential equations.
This work builds on our previous works on the topic. Order reduction ideas based on full rank decompositions were first introduced in 
\cite{robertezra} and \cite{robertezramartin}. These results were developed further in \cite{Sajjaal2013} and \cite{Sajjaal2019}. The present paper 
extends our prior work fundamentally in a number of ways. In the original work, one could (in one reduction step) only reduce a system to one whose
 index was  one less than  the index of the original  system; here one can reduce all the way to an index zero system (standard system) in one step.
Second, systems with inputs are considered. Third, missing links to established and classical descriptor results are established, revealing the utility of 
the approach advocated here. Finally, new reduced order forms are also introduced that are not considered in these previous papers.\newline

 Specifically our contributions may be summarized as follows.
 \begin{itemize}
 \item[(a)] We  consider first systems with no input.
It is known that, subject to some constraints, such a system can be  equivalently represented by a lower order  standard system.
 Since the order of a standard system cannot be reduced, this is the lowest order that can be achieved for the original descriptor system.
 There are situations where it is advantageous to obtain an equivalent  system description of lower  order {\em  but not necessarily  of minimal lower order.}
 This occurs, for example in analyzing switching linear descriptor systems \cite{Sajjaal2013,Sajjaal2019}.
Our first set of results is to demonstrate  how one can readily obtain various  equivalent  system descriptions of lower  order
for a linear descriptor system. 
\item[(b)]We also  give a simple procedure to reduce  a descriptor  system to an equivalent standard system.
\end{itemize}
Note that, although there are many results in the literature for reducing a descriptor system to a standard system (see \cite{Luenberger1977}, for one of the earliest results) there are very few results on
reducing to a lower order descriptor system, with the notable exception of \cite{Sajjaal2013}, and the results therein reduce the index of the system by one. 
The results in this present paper allow one to reduce a descriptor system to a lower order system of any lower index.

\begin{itemize}
\item[(c)]In the second part of our paper we consider systems with inputs and obtain two {coupled} reduced order systems associated with the original system in descriptor form.
These two systems lead directly to the celebrated quasi-Weierstrass form \cite{BergerIlchetAl2012} of the original system, {but in an elementary manner when compared with 
existing literature}. Recall the quasi-Weierstrass form gives rise a form that consists of two  subsystems which together are equivalent to the original system. 
One of these subsystems is a standard system whereas the other is very special type of descriptor system called a {\em pure descriptor system.}  As stated our derivation provides a simple way of constructing a quasi-Weierstrass form for a linear descriptor system, and relates our approach to existing mathematical results on Descriptor systems.
\end{itemize}

Our paper is structured as follows. We present preliminary material in Section 2. Our main results are derived in Sections 3 and 4. Examples illustrating the utility of our results are also given on Section 4.

\section{Preamble - Descriptor systems}

\noindent Consider a  linear time invariant (LTI) system
described by
the differential algebraic equation (DAE)
\begin{equation}\label{adescrip}
E\dot{x}=Ax
\end{equation}
where $x(t) \in \mathbb{C}^n$ is the system state at time $t \in \mathbb{R}$ and
 $ E, A \in \mathbb{C}^{n\times n}$. 
 When $E$ is nonsingular, this system is also described by the {\it standard system} $\dot{x} = E^{-1}Ax$.
If  $E$ is singular, then both algebraic equations and differential equations describe the behavior of the system, and the system is known 
as a {\em descriptor system.}\newline

 We say that system \eqref{adescrip} or $(E, \  A)$    is   {\it regular} if  the polynomial $\det(sE-A)$ is nonzero, that is, there exists $\lambda  \in \mathbb{C}$ such
 that $\lambda  E-A$ is nonsingular.
For such a scalar $\lambda $, we can   rewrite system \eqref{adescrip} as
 \[
 E\dot{x}=(A-\lambda  E)x + \lambda  Ex
 \]
and pre-multiply by  $(A-\lambda  E)^{-1}$ to obtain
 \begin{equation}
 \label{adescrip2}
 \F \dot{x} = (I +\lambda \F )x
 \end{equation}
 where
 \begin{equation}
 \label{eq:E0}
   \F  := (A-\lambda  E)^{-1}E
 \end{equation}
 We will find   this system description  useful  for several purposes, in particular for  reducing  system  \eqref{adescrip} to a system of lower order, that is, lower state dimension.\newline

 The  {\em  consistency space}  $\mathcal{C}= \mathcal{C}(E, A)$
for system (\ref{adescrip}) or $(E,A)$   is the set of all initial states $x_0\in \mathbb{C}^n$ for   which equation  (\ref{adescrip}) 
has a  classical (that is, differentiable) solution $x(\cdot):[0, \infty) \rightarrow \mathbb{C}^n$
with the  initial condition $x(0) = x_0$.
 We can characterize this with the following concept.
 The    {\em index} of a  matrix $F\in \mathbb{C}^{n\times n}$  
 is the smallest nonnegative integer $k^*$ for which $rank(F^{k^*+1}) = rank(F^{k^*})$
 where rank denotes the rank of a matrix;
this index is zero for a nonsingular matrix.
Note that the index of $F$ is also the smallest   nonnegative integer $k^*$ for which $\mathcal{R}(F^{k^*+1}) = \mathcal{R}(F^{k^*})$
where $\mathcal{R}$ denotes the image or range of a matrix.
Also $\mathcal{R}(F^k) = \mathcal{R}(F^{k*})$ for all $k\ge k^*$ and $ \mathcal{R}(F^{k}) \supset \mathcal{R}(F^{k*}) $ for $k\le k^*$.
If $F^{k^*} = 0$ we say that $F$ is {\em nilpotent}.

\begin{remark}
\label{rem:rem1}
It can readily be shown   that, for any $k=0,1,2, \dots$,  the subspace $\mathcal{R}(\F^{k})$  is the same 
for all $\lambda $ for which
$ \lambda  E-A$ is nonsingular \cite{CampbellBook1980}; hence the index of $\F$ is the same for
  all $\lambda $ for which
$ \lambda  E-A$ is nonsingular; we call this {\em the index of system (\ref{adescrip}) or $(E,A)$.}
It is also shown in  \cite{CampbellBook1980} that $\mathcal{C}(E, A) = \mathcal{R}(\F ^k)$ for $k\ge k^*$ where $k^*$ is the index of $\F $
and for  all $\lambda $ for which $\lambda  E-A $ is nonsingular.
\end{remark}
\begin{remark}
Since
$ \mathcal{R}   (\F ^{k^*+1})   =  \mathcal{R}( \F ^{k^*} ) =\mathcal{C}$ 
we see that $\F \mathcal{C} = \mathcal{C}$.
This implies that $\F $ is a one-to-one mapping of $\mathcal{C}$ onto itself; hence the kernel of $\F $
and $\mathcal{C}$ intersect only at zero.
Note that $\mathcal{C}=\{0\}$ if and only if $F$ is nilpotent;
in this case  we say that the system is  a {\em  pure descriptor system } and the only differentiable  solution is the zero solution $x(t) \equiv 0$.
If $\mathcal{C} \neq \{0\}$,  we let $\G$ be the inverse of the map $\F $ restricted to
$\mathcal{C}$,
that is, $\G\F x = x$  and  $\F \G x= x$ when $x \in \mathcal{C}$.
When the solution $x(t)$   
is in $\mathcal{C}$ for all $t$  then so is $\dot{x}(t)$; hence
multiplying \eqref{adescrip2} by $\G$ results in
\begin{equation}
\label{eq:descriptor3}
\dot{x} = \hA x 
\end{equation}
where $\hA = \G + \lambda  I$.
Also multiplying \eqref{eq:descriptor3} by $\F $ results in \eqref{adescrip2}.
Thus \eqref{eq:descriptor3} is equivalent to  \eqref{adescrip2};
hence \eqref{eq:descriptor3} and  \eqref{adescrip} are equivalent.
Thus the  restriction of the descriptor system to its consistency space is equivalent to the standard system (\ref{eq:descriptor3}) where $x(t)$ is in $\mathcal{C}$.
\end{remark}

\section{Reducing  a descriptor system}

Our first main result, Lemma \ref{th:main1}, shows how to simply reduce system \eqref{adescrip} to an equivalent  system of lower order and lower index.
It requires the following concepts and lemmas.
For a full column rank matrix $X$,
the matrix $X^{\dagger}$ denotes any {\em  left-inverse} of $X$, that is, it satisfies 
\[
X^{\dagger}X = I
\]
where $I$ is an identity matrix.
For example, $X^{\dagger} =  (X'X)^{-1}X'$.
We need the  following result  for an arbitrary $n \times n$ matrix $F$.\newline

\begin{lemma}
\label{lem:equiv}
Suppose $F\in \mathbb{C}^{n\times n}$,  $F^k \neq 0$ for some integer $k\ge 1$ and $X$ is a matrix of full column rank whose range equals that of $F^k$.
Then, for  any integer $l\ge 0$,
\begin{equation}
\label{eq:Prop1}
F^lX = X \tilde{F}^l			\qquad \mbox{where} \qquad \tilde{F} =X^{\dagger}FX\newline
\end{equation}
\end{lemma}
{\em Proof.}
Clearly it holds for $l=0$.
We now  prove, by induction that is holds for any $l\ge 1$.
We first show that \eqref{eq:Prop1} holds for $l=1$, that is,  $FX = X \tilde{F}$.
By assumption,
$\mathcal{R}(X) = \mathcal{R}(F^k)$; thus
\[
\mathcal{R}(FX) = \mathcal{R}(F^{k+1})\subset \mathcal{R}(F^k) = \mathcal{R}(X)
\]
that is $\mathcal{R}(FX) \subset \mathcal{R}(X)$.
So $FX = X\tilde{F}$  for some matrix $\tilde{F}$.
Multiplying both sides of this equation by any left-inverse $X^{\dagger}$ of $X$ yields $\tilde{F}=X^{\dagger}FX$.
Now suppose that  \eqref{eq:Prop1} holds for some integer $l \ge 1$. Then
\[
F^{l\!+\!1}X = FF^{l}X =FX \tilde{F}^{l}= X\tilde{F}\tilde{F}^{l} =X\tilde{F}^{l\!+\!1}
\]
Thus, \eqref{eq:Prop1} holds with $l$ replaced with $l+1$. By induction, it holds for all $l\ge 1$.
QED\newline



The following decomposition is useful in some of the  results of this paper.
Consider any  non-zero matrix $M\in \mathbb{C}^{n\times n}$.
A pair  of matrices $(X, Y)$   is  a {\em  full rank decomposition}  of $M$ if $X$ and $Y$ have maximum column rank and
 \begin{equation}
M=XY'
\end{equation}
If $r$ is the rank of $M$ then $r\le n$ and   $X,Y\in \mathbb{C}^{n\times r}$.
Clearly, $X$ and $M$ have the same range while $Y$ and $M'$ have the same range.
Also,
\begin{equation}
X=MY^ {\dagger'} \qquad \mbox{and} \qquad  Y=M'X^{\dagger'}
\end{equation}

\begin{lemma}
\label{Lemma2}
Suppose $F\in \mathbb{C}^{n\times n}$,  $F^k \neq 0$ for some integer $k\ge 1$ and $X,Y$ is a  full rank column rank decomposition of $F^k$.
Then,
\begin{equation}
X^{\dagger}FX = Y'FY^{\dagger\prime} =:\tilde{F}
\end{equation}
and  for  any integer $l\ge 0$,
\begin{equation}
F^{l+k} = X\tilde{F}^l Y'
\end{equation}
\end{lemma}

{\em Proof.}
Since $(X,Y)$ is a  full rank column rank decomposition of $F^k$, 
\begin{equation}
\label{eq:XY'}
F^k = XY'
\end{equation}
where $X,Y$  are  full column rank matrices
and the range of $X$ equals that of $F^k$.
Thus
$X=\F ^k Y^{\dagger '}$
and
\[
X^{\dagger}\F X=X^{\dagger}\F \F ^kY^{\dagger '} = X^{\dagger}\F ^k\F Y^{\dagger '}  =X^{\dagger}XY'\F Y^{\dagger '} = Y'\F Y^{\dagger '} 
\]

Consider any  integer $l\ge 0$.  
According to Lemma \ref{lem:equiv},  $ X\tilde{F}^l =F^lX $;
hence
\begin{equation}
\label{eq:Ftl}
\tilde{F}^l = X^{\dagger}F^lX
\end{equation}

Post-multiplying both sides of \eqref{eq:Ftl} by $Y'$ and using \eqref{eq:XY'}:
\begin{equation}
\label{eq:Ftl Flk}
\tilde{F}^lY' = X^{\dagger}F^lXY' = X^{\dagger}F^lF^k = X^{\dagger}F^{l+k}
\end{equation}
Since $\mathcal{R}(F^{l+k} )\subset \mathcal{R}(F^k) = \mathcal{R}(X)$, there exists a matrix $Y_l$ such that
\begin{equation}
\label{eq:Flk}
F^{l+k} = XY'_l
\end{equation}
 hence $X^{\dagger}F^{l+k} =X^{\dagger}XY'_l = Y_l'$.
 It now follows from \eqref{eq:Ftl Flk} that
$
Y'_l =\tilde{F}^lY' 
$
Combining this with \eqref{eq:Flk}  yields the desired result,
$
F^{l+k} = X\tilde{F}^l Y'
$.
QED\newline

We now obtain our first  reduction result.\newline

\begin{lemma}
\label{th:main1}
Consider a  regular  descriptor system described by
\eqref{adescrip}
and  any $\lambda  \in \mathbb{C}$ for which  $\lambda  E-A$ nonsingular.
For any integer $k\ge1$  with $F^k \neq 0$, where $\F$  is given by \eqref{eq:E0}, let
$X$ be  any  matrix of full column rank whose range equals that of $F^k$.
Then, 
$x(\cdot)$ is a differentiable solution to \eqref{adescrip} 
 if 
and  only if 
\begin{equation}
x = Xz
   \end{equation}
   and $z(\cdot)$ is a differentiable solution to
\begin{equation}
\label{eq:DAE4z}
\tF \dot{z} = (I+\lambda \tF ) z
\end{equation}
where
\begin{equation}
\label{eq:Etilde2}
\tF := X^{\dagger}\F X 
\end{equation}
Moreover $z=X^{\dagger}x$ and the index of \eqref{eq:DAE4z} is $\max \{ k^*\!-\!k, 0\} $ where $k^*$ is the index of \eqref{adescrip}  \newline
\end{lemma}

{\em Proof.}
When $x(\cdot)$ is a differentiable solution to \eqref{adescrip} we have $x(t) \in \mathcal{C}$ where $\mathcal{C}$ is the consistency space of $(E, A)$.
Since $\mathcal{C} \subset \mathcal{R}(F^k)$  
it follows that  $\mathcal{C}\subset  \mathcal{R}(X)$.
Hence, 
$
x = Xz
$
 and $z$ is uniquely given by
$
z= X^{\dagger}x
$.
As shown earlier, $x(\cdot)$ is a differentiable solution to \eqref{adescrip} 
if and only if it a solution of \eqref{adescrip2}
which is equivalent to
\begin{equation}
\label{eq:DAE5b}
\F X\dot{z} = (I+\lambda \F )Xz
\end{equation}
It follows from Lemma \ref{lem:equiv} that
$
 \F X = X\tF 
$
where $\tF$ is given by \eqref{eq:Etilde2}.
Thus \eqref{eq:DAE4z} is equivalent to
\begin{equation}
\label{eq:DAE6}
X\tF \dot{z} = X(I+\lambda \tF )z
\end{equation}
  Since $X$ has maximum column rank, 
  \eqref{eq:DAE6} is equivalent to 
 \eqref{eq:DAE4z}.
 To obtain the index of \eqref{eq:DAE4z}, choose any matrix $Y$ such that $(X,Y)$ is a full rank decomposition of $F^k$.
Recall from Lemma \ref{Lemma2} that for any $l\ge 0$,
 $F^{l+k} = X\tilde{F}^l Y'$.
 Since $X$ has maximum column rank the matrices $F^{l+k}$ and $\tilde{F}^l Y'$ have the same rank.
  Since $Y'$ has maximum row rank the matrices   $\tilde{F}^l Y'$ and $\tilde{F}^l$ have the same range;
  hence $F^{l+k}$ and $\tilde{F}^l$ have the same rank.
  It now follows that if $k\le k^*$ then the  index $l^*$ of \eqref{eq:DAE4z} is $k^*-k$ and if $k >k^*$ we have $l^* = 0$.
QED

\begin{remark}
For a descriptor system with singular $E$, the rank of the matrix $\F$ is less than $n$; thus  the rank of 
$\F^k$ and, hence,  $X$ is less than $n$.
Since $X$ has maximum column rank this tells us that the state $z$ of the new system in \eqref{eq:DAE5b} 
is in $\mathbb{C}^m$ with $m<n$. Hence  \eqref{eq:DAE4z}  is an equivalent reduced order version of the original system \eqref{adescrip}.
\end{remark}

\begin{example}
\label{ex1}
{\rm
To illustrate Lemma \ref{th:main1},
consider a descriptor system described  by \eqref{adescrip} with 
\[
E = 
\left(\begin{array}{rrr}
2	&-2	&-2	\\
2	&2	& -2	\\
0	&0	&0
\end{array}
\right)
\,,
\qquad
A =
\left(\begin{array}{rrr}
1 &   1   &  1	\\
1	&-1	& 1	\\
1   &1 &   -1
     \end{array}
\right)
\]
Since $A$ is non-singular, we can consider $\lambda = 0$ ; hence 
\[
F =A^{-1}E =
\left(\begin{array}{rrr}
1 &  1 & -1	\\
0 &  -2  & 0	\\
    1 &  -1 &  -1
     \end{array}
\right)
\]
The rank of $F$ is two whereas that of 
\begin{equation}
\label{eq:F^2}
F^2 =
\left(\begin{array}{rrr}
0 &  0 & 0 	\\
0 &  4  & 0	\\
    0&  4 &0
     \end{array}
\right)
\end{equation}
 and $F^3$ is one.
Thus this is an index two system whose consistency space is the range of $F^2$. Considering $k=1$, the  full column rank matrix 
\[
X = \left(\begin{array}{rr}
1&  1	\\
   0	&  -2\\
   1  & -1
     \end{array}
     \right)
\]
has the  same  range as that of $F$.
Hence this system can be described by $x=Xz$ and $\tilde{F} \dot{z}= z$ where $z= X^{\dagger}x$ and
\[
\tilde{F} = X^{\dagger}FX =
 \left(\begin{array}{rr}
0 & 2	\\
0 & -2
    \end{array}
    \right)
\]
which is an index  one matrix. Considering $k=2$,  the range of full column rank matrix 
\begin{equation}
\label{eq:X}
X = \left(\begin{array}{r}
  0	\\
   1	\\
   1
     \end{array}
     \right)
\end{equation}
is the same  as that of $F^2$ and is the consistency space.
Here
$
\tilde{F} = X^{\dagger}FX =
-2
$.
Hence the original descriptor system can be described by the standard system
\[
-2\dot{z} =z
\]
and $x=Xz= [0 \ z \ z]^T$. Also $z=X^{\dagger}x = (x_1 + x_2)/2$.\newline
}
\end{example}

We now obtain the following result  for an arbitrary $n \times n $ matrix $F$.
This shall be used to obtain another reduction result; namely, Lemma \ref{lem:Y'}.\newline


\begin{lemma}
\label{lem:Y'x}
Suppose  that $F\in \mathbb{C}^{n \times n}$ has index $k^*$
 and $Y$ is a   matrix whose range is the same as that of
 $F^{'k}$
for  some integer $k\ge 1$.
Then, $Y'F^lx \neq 0$  for all nonzero $x\in \mathcal{R}(F^{k^*})$ and all nonnegative integers $l$.\newline
\end{lemma}

\noindent
{\em Proof.}
Consider any nonnegative integer $l$. Suppose  that $Y'F^lx=0$ for some $x\in \mathcal{R}(F^{k^*})$. 
Since the range of $Y$ is the same as that of $F^{'k}$,
$F^{'k} = Y\hat{X}'$   for some  matrix $\hat{X}$ and $F^k= \hat{X}Y'$.
Hence,
\begin{equation}
\label{eq:0=Y'x}
0=\hat{X}Y'F^lx = F^kF^lx = F^{k+l}x
\end{equation}
Since
 $F$ has index $k^*$,
\[
\mathcal{R}(F^{k^*}) =    \mathcal{R}(F^{k+l+k^*}) = F^{k+l} \mathcal{R}(F^{k^*})
 \]
thus,  $F^{k+l} \mathcal{R}(F^{k^*})=  \mathcal{R}(F^{k^*})$.
This implies that $F^{k+l}$ is a one-to-one mapping of 
$\mathcal{R}(F^{k^*})$ onto itself; hence the kernel of $F^{k+l}$
and $\mathcal{R}(F^{k^*})$ intersect only at zero.
Now \eqref{eq:0=Y'x} implies that 
 that $x=0$.\newline

We now obtain a second  reduction result.\newline

\vspace{0em}
\begin{lemma}
\label{lem:Y'}
Consider a  regular  descriptor system described by
\eqref{adescrip}
and  any $\lambda  \in \mathbb{C}$ for which  $\lambda  E-A$ nonsingular.
For any integer $k\ge1$  with $F^k \neq 0$, where $\F$  is given by \eqref{eq:E0}, let
$Y$ be any matrix of maximum column rank whose range is the same as that of    $\F^{'k}$.   %
Then, there is a matrix $H$ such that
$x(\cdot)$ is a differentiable solution to \eqref{adescrip} 
if and   only if  
\begin{equation}
\label{eq:H}
x = Hz
   \end{equation}
   and $z(\cdot)$ is a differentiable solution to
\begin{equation}
\label{eq:DAE4z0}
\tF \dot{z} = (I+\lambda \tF ) z
\end{equation}
where
\begin{equation}
\label{eq:Etilde3}
\tF = Y'\F Y^{\dagger'}
\end{equation}
Moreover 
\begin{equation}
\label{eq:z=Y'x1}
z=Y'x
\end{equation}
and the index of \eqref{eq:DAE4z0} is $\max \{ k^*\!-\!k, 0\} $ where $k^*$ is the index of \eqref{adescrip}.\newline
\end{lemma}

\noindent
{\em Proof.}
As shown earlier, $x(\cdot)$ is a differentiable solution to \eqref{adescrip} 
if and only if it is a solution of \eqref{adescrip2}. 
Introducing $\hat{x} = F^kx$ we obtain that
\begin{equation}
\label{eq:hat}
F\dot{\hat x} =(I+ \lambda F)\hat{x}
\end{equation}
Using Lemma \ref{th:main1}, 
$\hat{x}(\cdot)$ is a differentiable solution to \eqref{eq:hat} 
 if 
and  only if 
\begin{equation}
\hat{x} = Xz
   \end{equation}
   and $z(\cdot)$ is a differentiable solution to
\begin{equation}
\label{eq:DAE4zhat}
\tF \dot{z} = (I+\lambda \tF ) z
\end{equation}
where
\[
\tF = X^{\dagger}\F X 
= Y'\F Y^{\dagger'}
\]
The second equality comes from Lemma \ref{Lemma2}.
The index of \eqref{eq:DAE4zhat} is $\max \{ k^*\!-\!k, 0\} $ where $k^*$ is the index of \eqref{adescrip} and
\[
 z=X^{\dagger}\hat{x} = X^{\dagger}F^k x= X^{\dagger}XY'x = Y'x
 \]
Lemma \ref{lem:Y'x} tells us that the kernel  of $Y'$ and  $\mathcal{C}$
intersect only at zero, there is a unique matrix $H$ such that \eqref{eq:H} holds.
QED\newline
\vspace{0em}

\begin{example}
{\rm
To illustrate Lemma \ref{lem:Y'},
recall the system in Example \ref{ex1}. We see that
\[
Y= \left(\begin{array}{rr}
1&  0	\\
   1	&  1\\
   -1  & 0
     \end{array}
     \right)
\]
is a full column rank matrix whose range is the same as that of $F'$.
Hence this system can be described by
$\tilde{F} \dot{z}= z$ where $z= Y'x$ and
\[
\tilde{F} =Y'FY^{\dagger \prime}=
 \left(\begin{array}{rr}
0 & 0	\\
0 & -2
    \end{array}
    \right)
\]
which is a index one matrix. Since $z_2 = x_2$ and $x$ must be in the range of the matrix $X$  in \eqref{eq:X} (the consistency space), we must have
$x= [ 0\  z_2 \  z_2 ]^T$. Considering $k=2$  the full column rank matrix 
\[
Y= \left(\begin{array}{r}
  0	\\
   1	\\
   0
     \end{array}
     \right)
\]
has the same  range  as that of $F'^2$.
Here
$
\tilde{F} = Y'FY^{\dagger\prime} =
-2
$.
Hence the original descriptor system can be described by the standard system
$
-2\dot{z} =z
$
and  $z=Y'x =  x_2$. 
Since $x_2 =  z$ and $x$ must be in the range of the matrix $X$ in \eqref{eq:X}, we must have
$x= [ 0\  z \  z ]^T$.

}
\end{example}

\begin{remark}
Suppose that $(X, Y)$ is a full rank decomposition of the matrix $\F$ in \eqref{eq:E0}.
Then  $\F = XY'$.
Considering the result in Lemma \ref{th:main1} for $k=1$,
we see that the matrix $\tF$ in \eqref{eq:Etilde2} is given by
\[
\tF = X^{\dagger}FX = X^{\dagger}XY'X = Y'X
\]
This along with  Lemma \ref{th:main1}  and and $\lambda = 0$ captures a corresponding result in \cite{Sajjaal2013} when  $A$ is nonsingular.
\end{remark}

\subsection*{Application to switching linear systems}
The above results can be useful in reducing a switching descriptor system to a lower order system.
To illustrate,
consider a switching descriptor system described by
\begin{equation}
\label{eq:switchSys}
E_{\sigma(t)} \dot{x} = A_{\sigma(t)} x
\end{equation}
where $\sigma(t) \in \left\{1,2, \ldots, N\right\}$ and $E_i, A_i \in \mathbb{C}^{n \times n}$ for $i=1,2, \dots, N$.
Suppose that for some $\lambda \in \mathbb{C}$ and for each $i$ there exists $k_i$ such that
{\em the range of $F_i^{k_i}$ is the same for all $i$} where $F_i = (A_i-\lambda E_i)^{-1}E_i$.
Recalling Lemma \ref{th:main1}, let 
$X$ be any matrix of maximum column rank whose range is the same as that of  $\F_i ^{k_i}$ for all $i$ .
Then, 
$x(\cdot)$ is a differentiable solution to \eqref{eq:switchSys} 
 if 
and  only if 
$
x = Xz
$
   and $z(\cdot)$ is a differentiable solution to the lower order switching system
\begin{equation}
\label{eq:switchSysLow}
\tF_{\sigma(t)} \dot{z} = (I+\lambda \tF_{\sigma(t)}) z
\end{equation}
where
$
\tF_i := X^{\dagger}\F_i X 
$.
Moreover $z=X^{\dagger}x$.

\section{Equivalent standard systems}
We have already seen that  \eqref{adescrip} is equivalent to a standard system on the consistency space.
Here we  provide simple characterizations of  reduced order  standard systems which are equivalent to \eqref{adescrip}.
Lemma \ref{th:main1}  leads to the following result which yields an equivalent lower order standard system for the 
original descriptor system \eqref{adescrip}.\newline

\begin{corollary}
\label{th:main}
Consider a  regular non-pure  descriptor system described by
\eqref{adescrip}
and  any $\lambda  \in \mathbb{C}$  for which  $A-\lambda  E$ nonsingular.
With $\F $ given by \eqref{eq:E0}
  let
 $X$ be any  full column rank matrix whose range  is the same as that of
$\F^k$     for some integer 
    $k\ge k^*$ where $k^*$ is the  index of $(E, A)$.
Then
$X^{\dagger}\F X$
is  nonsingular and
$x(\cdot)$ is a differentiable solution to \eqref{adescrip} 
 if 
and  only if 
\begin{equation}
\label{eq:xAndZ}
x = Xz
   \end{equation}
   and $z(\cdot)$ is a differentiable solution to
\begin{equation}
\label{eq:DAE4}
\dot{z} = \tA z
\end{equation}
where
\begin{equation}
\label{eq:Atilde}
\tA= (X^{\dagger}\F X)^{-1} + \lambda I
\end{equation}
Moreover
\begin{equation}
\label{eq:z=Xdagx}
z=X^\dagger x\newline
\end{equation}
\end{corollary}

\vspace{1em}

When $A$ is invertible, one can choose $\lambda =0$. In this case, we obtain the following simpler expressions:
\begin{equation}
F=A^{-1}E, \qquad \tilde{A}= (X^{\dagger}A^{-1}EX)^{-1}\newline
\end{equation}

 Lemma \ref{lem:Y'}  leads to the following result which yields another  equivalent lower order standard system for the 
original descriptor system \eqref{adescrip}.\newline

\begin{corollary}
\label{lem:standard}
Consider a  regular non-pure descriptor system described by
\eqref{adescrip}
and  any $\lambda  \in \mathbb{C}$ for which  $\lambda  E-A$ nonsingular.
With $\F $ given by \eqref{eq:E0},
let
$Y$ be any matrix of maximum column rank whose range is the same as that of    $\F^{'k}$  
   for some
  integer   $k\ge k^*$ where $k^*$ is the  index of $(E, A)$.
Then
$Y'FY^{\dagger'}$
is  nonsingular and
$x(\cdot)$ is a differentiable solution to \eqref{adescrip} 
if and   only if  
\begin{equation}
\label{eq:H2}
x= Y^{\dagger'}z
   \end{equation}
   and $z(\cdot)$ is a differentiable solution to
\begin{equation}
\label{eq:DAE5}
\dot{z} = \tA z
\end{equation}
where
\begin{equation}
\label{eq:AtildeY}
\tA= (Y'\F Y^{\dagger'})^{-1} + \lambda I
\end{equation}
Moreover
\begin{equation}
\label{eq:z=Y'x2}
z=Y'x
\end{equation}

\end{corollary}
{\em Proof.}
We just need to show that $H= Y^{\dagger'}$.
Since $Y'x \neq 0$  holds for all $x$ in the consistency space $\mathcal{C}$ of \eqref{adescrip},
it follows that  $\left\{z: z=Y'x \mbox{ and } x\in \mathcal{C}\right\} =\mathcal{C}^m$ where $m$ equals the dimension of $\mathcal{C}$ and the number of columns of $Y$.
Using  \eqref{eq:z=Y'x1} and \eqref{eq:H} we now obtain that 
$z= Y'Hz$ for all $z\in\mathcal{C}^m$. Hence $Y'H = I$ from which it follows that $H=Y^{\dagger '}.$
QED

\vspace{1em}
When $A$ is invertible, one can choose $\lambda =0$. In this case, we have the simpler expressions:
\begin{equation}
F=A^{-1}E, \qquad \tilde{A}= (Y'A^{-1}EY^{\dagger'})^{-1} 
\end{equation}

The following result leads to further expressions for $\tilde{A}$.\newline
\begin{lemma}
\label{cor:nonsingular}
Suppose that $F\in \mathbb{C}^{n \times n}$ is a matrix which is not nilpotent, has index $k^*$
and $X$ and $Y$ are full column rank matrices whose ranges are  the same as that of
$F^k$ and $F^{'k}$, respectively, 
for  some integer $k\ge k^*$.
Then, $Y'F^lX$ is nonsingular for every nonnegative  integer $l$.\newline
\end{lemma}

\noindent
{\em Proof.}
Consider any nonnegative integer $l$ and suppose  that $Y'F^lXz=0$. 
Since the vector $Xz$ is in $\mathcal{R}(F^{k})$ and $k\ge k^*$,  this vector is in $\mathcal{R}(F^{k^*})$.
It now  follows from
Lemma \ref{lem:Y'x} that $Xz=0$.
 Since $X$ has maximum column rank we obtain that $z$ is zero.
 With $Y$  and $X$ having  the same dimensions,  $Y'F^lX$ is square.
Thus $Y'F^lX$ is nonsingular.
QED

\begin{remark}
\label{rem:Y'X}
Consider a non-pure system described by \eqref{adescrip}. Then $F^k \ne 0$ for every nonnegative integer $k$  where  $\F$ is  given by \eqref{eq:E0}.
Suppose that $X$ and $Y$ are full column rank matrices whose ranges are the same as that of
$\F^k$ and $\F^{'k}$, respectively,    where  $k\ge k^*$ and $k^*$ is the index of $\F$.
 Then,
the above result tells us that $Y'X$ is invertible.
Since $(Y'X)^{-1}Y'X = I$,  a left-inverse of $X$ is given by
\begin{equation}
\label{eq:Xdag}
X^{\dagger} = (Y'X)^{-1}Y'
\end{equation}
Hence
\begin{equation}
\label{eq:Xdag}
X^{\dagger}\F X = (Y'X)^{-1}Y'\F X
\end{equation}
and the matrix in \eqref{eq:DAE4} is given by
\begin{equation}
\label{eq:tildeA}
\tilde{A} = (Y'\F X)^{-1}Y'X + \lambda I
\end{equation}
Since,
$(Y'X)^{-'}X'Y = I$,  a left-inverse of $Y$ is given by
\begin{equation}
\label{eq:Ydag}
Y^{\dagger} = (Y'X)^{-'}X'
\end{equation}
Hence $Y'\F Y^{\dagger'} = Y'\F X(Y'X)^{-1}$
and the matrix in \eqref{eq:DAE5} is given by
\begin{equation}
\label{eq:tildeE2Y}
\tilde{A} = Y'X(Y'\F X)^{-1} + \lambda I
\end{equation}
\end{remark}

\paragraph*{An equivalent full order standard system on the consistency space}

Using the results in Corollary \ref{th:main} or Corollary \ref{lem:standard} we can obtain a standard 
system which is equivalent to the original descriptor system and has the same state as the original system.

\begin{lemma}
Consider a non-pure system described by \eqref{adescrip}.
Suppose that $X$ and $Y$ are full column rank matrices whose ranges are the same as that of
$\F^k$ and $\F^{'k}$, respectively,    where  $k\ge k^*$ and $k^*$ is the index of $\F$.
Then, $Y'X$ and $Y'FX$ are  nonsingular and
$x(\cdot)$ is a differentiable solution to \eqref{adescrip} 
if and   only if  $x(t)$ is in the range of $X$ and
\begin{equation}
\label{eq:descriptor3a}
\dot{x} = \hat{A}x
   \end{equation}
   where
   \begin{equation}
\hA = X(Y'\F X)^{-1}Y' + \lambda X(Y'X)^{-1}Y'	
\end{equation}
\end{lemma}

{\em Proof.}
Lemma \ref{cor:nonsingular} tells us that $Y'X$ and $Y'FX$ are nonsingular.
It follows from \eqref{eq:xAndZ}, \eqref{eq:DAE4} and \eqref{eq:z=Xdagx}    that the behavior of $x$ is described by \eqref{eq:descriptor3a} with
$\hA =X\tA X^\dagger $.
Recalling \eqref{eq:tildeA} and \eqref{eq:Xdag}  we see that
\begin{align*}
\hA &= X(Y'FX )^{-1}(Y'X)(Y'X)^{-1}Y' + \lambda X(Y'X)^{-1}Y'		\\
&=X(Y'FX )^{-1}Y' + \lambda X(Y'X)^{-1}Y'	
\end{align*}
One obtains the same result using \eqref{eq:H2}, \eqref{eq:DAE5} and \eqref{eq:z=Y'x2}   along with  \eqref{eq:Ydag} and \eqref{eq:tildeE2Y}.
QED\newline

When $E$ is invertible,  consider any $\lambda$ for which $A-\lambda E$ is invertible.
In this case the index $k^*$ of $F=(A-\lambda E)^{-1}E$ is zero.
Hence $X$ and $Y$ are invertible one can readily show that $\hat{A} = E^{-1}A$.
When $A$ is invertible, one can choose $\lambda =0$. In this case,  $F=A^{-1}E$ and 
\begin{equation}
 \hat{A}=X(Y'A^{-1}EX )^{-1}Y'
\end{equation}

\section{Systems with inputs}

We now consider systems with inputs described by
\begin{equation}
\label{eq:sysinput}
E\dot{x} = Ax + Bu
\end{equation}
where $u(t) \in \mathbb{C}^m$ is the system input and $B\in \mathbb{C}^{n\times m}$.
When $u=0$, a classical solution to \eqref{eq:sysinput} is constrained to the consistency space associated with
\eqref{eq:sysinput}. When $u\neq0$ this is not necessarily the case and we need further analysis.
When $(E, A)$ is regular, there exists $\lambda \in \mathbb{C}$ such that $A-\lambda E$ is nonsingular and,
following the derivation of $\eqref{adescrip2}$, we see that
\eqref{eq:sysinput} 
is equivalent to
\begin{equation}
\label{eq:sysinput2}
\F\dot{x} = (I+\lambda \F)x + \G u
\end{equation}
where $\F$ is given by \eqref{eq:E0} and
\begin{equation}
 \label{eq:G}
\G:=(A-\lambda E)^{-1}B\newline
\end{equation}

Using the following corollary to Lemma \ref{lem:equiv} we  can obtain our first  result, Lemma \ref{lem:Yu'}.\newline

 \begin{corollary}
\label{cor:equiv}
Suppose $F\in \mathbb{C}^{n\times n}$,  $F^k \neq 0$ for some integer $k\ge 1$  and $Y$ is a matrix of full column rank whose range equals that of $F^{'k}$.
Then, for  any integer $l\ge 1$,
\begin{equation}
\label{eq:Prop2}
Y'F^l=  \tilde{F}^lY'	\qquad \mbox{where} \qquad \tilde{F} =Y'FY^{\dagger '}\newline
\end{equation}
for $l=1,2, \dots$ where $\tilde{F} =Y'FY^{\dagger '}$.\newline
\end{corollary}

\begin{lemma}
\label{lem:Yu'}
Consider a  regular  descriptor system described by
\eqref{eq:sysinput}
and  any $\lambda  \in \mathbb{C}$ for which  $\lambda  E-A$ nonsingular.
For any integer $k\ge1$  with $F^k \neq 0$, where $\F$  is given by \eqref{eq:E0}, let
$Y$ be any matrix of maximum column rank whose range is the same as that of    $\F^{'k}$.   %
Suppose 
$x(\cdot)$ is any differentiable solution to \eqref{eq:sysinput} 
and let
\begin{equation}
\label{eq:z=W'x1}
z_1=Y'x
\end{equation}
 Then $z_1(\cdot)$ is a differentiable solution to
\begin{equation}
\label{eq:DAE4z2}
\tF \dot{z}_1 =  (I+\lambda\tF)z_1 +\tGone u
\end{equation}
where
\begin{align}
\tF =  Y'\F  Y^{\dagger'},\qquad 
\tGone= Y'\G
\end{align}
\end{lemma}

\noindent
{\em Proof.}
As shown above, $x(\cdot)$ is a differentiable solution to \eqref{eq:sysinput} 
if and only if it a solution to \eqref{eq:sysinput2}. 
Hence
\[
Y'\F \dot{x} =Y'(I+\lambda \F)x +Y'\G u
\]
 Corollary \ref{cor:equiv} tells us that $Y'\F = \tF Y'$ where $\tF = Y'\F Y^{\dagger '}$; hence
 \[
\tF \dot{z}_1 = (I+\lambda \tF )z_1 +\tGone u
 \]
 where  $z_1=Y'x$.
QED
\vspace{1em}

\begin{remark}
\label{rem:standard}
 If  $k\ge k^*$ in the above lemma, where $k^*$ is the index of $(E,A)$
then,  
$ Y'F Y^{\dagger'}$ is nonsingular; hence
\eqref{eq:DAE4z2} is equivalent to  the standard system
\begin{equation}
\label{eq:}
 \dot{z}_1 =  \tA  z_1 +\tBone u
\end{equation}
where 
\begin{equation}
\label{eq:At}
\tA = (Y'\F Y^{\dagger'})^{-1} +\lambda I,\qquad \tBone =(Y'\F Y^{\dagger'} )^{-1}Y'\G
\end{equation}
\end{remark}

With a nonzero input $u$, the state $x$ is not confined to the consistency space and we cannot recover $x$ from $z_1$.
So, now we proceed to obtain another reduced order system which contains further information on $x$.
To achieve this, need the following result for an arbitrary square matrix $F$; this result  is analagous to Lemma \ref{lem:equiv}.\newline

\begin{lemma}
\label{lem:VFV}
Suppose that    $F\in \mathbb{C}^{n \times n}$  is singular 
and  $V$ is any matrix of maximum column  rank whose range equals the  kernel of $F^k$ for some  integer $k\ge 1$.
Then, for  any integer $l\ge 1$,
\begin{equation}
\label{eq:ninpotent1}
F^lV = VN^l
\end{equation}
 where 
 $N =V^{\dagger}FV$.
 Moreover $N^k=0$.\newline
\end{lemma}

\noindent
{\em Proof.}
We prove this by induction.
We first show that \eqref{eq:ninpotent1}   holds for $l=1$, that is,
$
FV = VN
$.
If $v$ is  the range of $V$, then   $F^kv =  0$.
Thus $ F^k(Fv) = F(F^{k}v)= 0$;   this implies that $Fv$ is in the kernel of $F^k$ and, hence, it is in the range of $V$.
Thus $\mathcal{R}(FV) \subset \mathcal{R}(V)$.
This means that
 $FV= VN$ for some matrix $N$.
Multiplying both sides of this equation by $V^{\dagger}$ results in
$N  =V^{\dagger}FV$.
Thus, \eqref{eq:ninpotent1}  holds for $l=1$. Now suppose that for some integer $l^* \ge 1$, \eqref{eq:ninpotent1} holds  with    $l =l^*$. Then
\[
F^{l^*\!+\!1}V = FF^{l^*}V = FVN^{l^*}=VNN^{l^*}= VN^{l^*\!+\!1}
\]
Thus \eqref{eq:ninpotent1} holds with $l=l^*+1$.
By induction, it holds for all $l\ge 1$. It follows from \eqref{eq:ninpotent1} that $F^kV=VN^k$; hence
 $N^k = V^{\dagger}F^kV$.
 Since   the  range of $V$ is the  kernel of $F^k$,
 $F^{k}V=0$; thus $N^{k} = 0$.
QED\newline

The following result is a simple corollary to Lemma \ref{lem:VFV}.\newline

\begin{corollary}
\label{cor:equiv2}
Suppose that    $F\in \mathbb{C}^{n \times n}$ is singular
and  $W$ is any matrix of maximum column  rank whose range equals the  kernel of $F'^k$ for some $k\ge 1$.
Then,  for any integer $l\ge 1$,
\begin{equation}
\label{eq:ninpotent2}
W'F^l =  N^lW'
\end{equation}
  where 
 $N =W'FW^{\dagger'}$.
 Moreover $N^k=0$.\newline
\end{corollary}

Using Corollary \ref{cor:equiv2} we obtain another reduced order subsystem associated 
with descriptor system \eqref{eq:sysinput}.

\begin{lemma}
\label{lem:W'}
Consider a  regular  descriptor system described by
\eqref{eq:sysinput}
and  any $\lambda  \in \mathbb{C}$ for which  $\lambda  E-A$ nonsingular.
For any  integer $k\ge1$ let
$W$ be any matrix of maximum column rank whose range is the same as that of  
the kernel of  $\F^{'k}$  with $\F $ given by \eqref{eq:E0}.   %
Suppose 
$x(\cdot)$ is a differentiable solution to \eqref{eq:sysinput}
and let
\begin{equation}
\label{eq:z=W'x1}
z_2=W'x
\end{equation}
 Then $z_2(\cdot)$ is a differentiable solution to
\begin{equation}
\label{eq:DAE4z3}
\tilde{N} \dot{z}_2=  z_2 +\tilde{B}_2u
\end{equation}
where
\begin{align}
\label{eq:ENtilde}
\tilde{N} = (I+\lambda W'F W^{\dagger'})^{-1}W'F W^{\dagger'},\,
\tilde{B}_2= (I+\lambda W'F W^{\dagger'})^{-1}W'G
\end{align}
and $\tilde{N}^k=0$.\newline
\end{lemma}

\noindent
{\em Proof.}
As shown earlier, $x(\cdot)$ is a differentiable solution to \eqref{eq:sysinput} 
if and only if it a solution to \eqref{eq:sysinput2}. 
Hence
\[
W'F\dot{x} =W'(I+\lambda F)x +W'Gu
\]
 From  Corollary \ref{cor:equiv2}, 
  $W'F = NW'$  where $N= W'FW^{\dagger'}$ and $N^k =0$; hence
 \[
 N\dot{z}_2 = (I+\lambda N)z_2 +W'Gu
 \]
 where  $z_2=W'x$.
 Since $N^k =0$ the eigenvalues of $N$ are zero;
 hence the eigenvalues of $I+\lambda N$ are one, so 
  $I+\lambda N$ is invertible and 
  we obtain the desired result that
  \[
   (I+\lambda N)^{-1} N\dot{z}_2 =  z_2   +(I+\lambda N)^{-1}W'Gu
   \]
   Since $(I+\lambda N)^{-1}$ and $N$ commute, $\tilde{N} =  (I+\lambda N)^{-1} N$ and $N^k=0$, it follows that
   $\tilde{N}^k =(I+\lambda N)^{-k}N^k = 0$.
QED
\vspace{1em}


\subsection{Quasi-Weierstrass form}
We have obtained  two subsystems \eqref{eq:DAE4z2} and \eqref{eq:DAE4z3}
associated with the original descriptor system \eqref{eq:sysinput}.
In order for these two subsystems to completly describe the behavior of the original system, we need the 
matrix $[Y \; W]$ to be nonsingular.  This turns out to be the case if we consider $k\ge k^*$, the index of the original system.
To prove this we first obtain  the following result for an arbitrary square matrix.\newline

\begin{lemma}
\label{lem:V'Wnonasing}
Suppose  $F\in \mathbb{C}^{n \times n}$  is singular,  is not nilpotent, has index $k^*$
and $V, W$ are any matrices  of maximum column  rank whose ranges are the kernels of $F^k$ and $F^{'k}$, respectively, for  some $k\ge k^*$.
Then 
$V'W$  is    nonsingular.\newline
\end{lemma}

{\em Proof.}
To show that $V'W$ is nonsingular, suppose $V'Wz=0$. Then $Wz$ is in the orthogonal complement of the range  of $V$ which equals the range of $F^{'k}$.
Hence $Wz = Y\xi$ for some vector $\xi$ where $Y$ is a full column rank matrix whose range equals that of $F^{'k}$.
Let $X$ be a full column rank matrix whose range equals that of $F^k$.
Then the range of $X$ equals the orthogonal complement of the range of $W$
and $ X'Y\xi =X'Wz=0$.
Lemma \ref{cor:nonsingular} tells us that $X' Y=(Y'X)'$ is nonsingular.
Thus $\xi$ is zero
  and since $W$ has maximum column rank, $z=0$.
This implies that $V'W$ is nonsingular.
QED

We can now prove that $T=[Y\; W]$ is invertible for $k\ge k^*$.

\begin{lemma}
Suppose  $F\in \mathbb{C}^{n \times n}$  is singular,  is not nilpotent and has index $k^*$.
For any $k\ge k^*$,  
 let $X$ and $Y$ be any matrices  of maximum column  rank whose ranges are the same as that of $F^k$ and $F^{'k}$, respectively, and 
let $V$ and $W$ be  any matrices  of maximum column  rank whose ranges are the kernels of $F^k$ and $F^{'k}$,
respectively, 
Then
$\left[ 
Y	\;
W
\right]$
 is   nonsingular 
with inverse
\begin{equation}
\left[ \begin{array}{c}
(X'Y)^{-1}X'\\
(V'W)^{-1}V'
\end{array}\right]\newline
\end{equation}

 \end{lemma}
 \noindent
 {\em Proof.}
 Since $k\ge k^*$,  where $k^*$ is the index of $F$, we know from Lemma \ref{cor:nonsingular} and Lemma  \ref{lem:V'Wnonasing} that $X'Y$ and
 $V'W$ are nonsingular.
Since the range of $W$ is the kernel of  $F^{'k}$ we have $F^{'k}W=0$; hence
 $W'F^{k} = 0$. 
  Since the range of $X$ is $F^k$, we must have $X'W =(W'X)'=0$.
 Using the same reasoning we also have $Y'V=0$.
 Hence
 \[
 \left[ \begin{array}{c}
(X'Y)^{-1}X'\\
(V'W)^{-1}V'
\end{array}\right]
\left[
\begin{array}{cc}
Y & W
\end{array}
\right]
=
 \left[ \begin{array}{cc}
(X'Y)^{-1}X'Y		& 	(X'Y)^{-1}X'W\\
(V'W)^{-1}V'Y				&	(V'W)^{-1}V'W
\end{array}\right] \]
\[
=
 \left[ \begin{array}{cc}
I	& 	0\\
0			&	I
\end{array}\right]
 \]
 QED\newline

 Using the above lemma along with Remark \ref{rem:standard} and Lemma \ref{lem:W'} we obtain 
 a decomposition of the original system into a standard system and a pure descriptor system.
This  decomposition is obtained in  \cite{BergerIlchetAl2012} and is referred to as a {\em quasi-Weierstrass form} of
\eqref{eq:sysinput}.
The derivation in \cite{BergerIlchetAl2012} is based on the Wong sequences presented in \cite{Wong1974}.
We believe the derivation here is more elementary. 
Also, one may simply compute the matrices involved here by performing a singular value decomposition
of $F^k$ where $k$ is greater than or equal to the index of $(E, A)$; see Remark \ref{rem:svd} below.\newline

\begin{theorem}
\label{th:qw}
Consider a    regular non-pure descriptor system of index $k^*$ described by
\eqref{eq:sysinput}  with $E$ singular
and  any $\lambda  \in \mathbb{C}$ for which  $\lambda  E-A$ is nonsingular.
With $\F $ given by \eqref{eq:E0} and 
for any integer $k\ge k^*$, 
 let $X$ and $Y$ be any matrices  of maximum column  rank whose ranges are the same as that of $F^k$ and $F^{'k}$, respectively,  and
let $V$ and $W$ be  any matrices  of maximum column  rank whose ranges equal the kernels of $F^k$ and $F^{'k}$,
respectively.
Then $x(\cdot)$ is a differentiable solution to \eqref{eq:sysinput}   if and only if
\begin{equation}
x= X(Y'X)^{-1}z_1 +V(W'V)^{-1}z_2
\end{equation}
and
\begin{align}
\dot{z}_1 = \tilde{A} z_1 +\tilde{B}_1u\\
\tilde{N}\dot{z}_2 = z_2 +\tilde{B}_2 u
\end{align}
where $\tilde{A}$ and $\tilde{B}_1$  are given by \eqref{eq:At}  while $\tilde{N}$ and $\tilde{B}_2$  are  given by \eqref{eq:ENtilde}.
Moreover
\[
 \tilde{N}^{k}=0
\]
and
\[
z_1 =Y'x,\qquad z_2 = W'x
\]
\end{theorem}

\begin{example}
{\rm
To illustrate Theorem \ref{th:qw}, consider descriptor  system \eqref{eq:sysinput} with $A$ and $E$ as given in Example \ref{ex1} and
\[
B=
\left[\begin{array}{r}
0\\0\\1
\end{array}
\right]
\]
Here $k^* = 2$ and $F^2$ is given in \eqref{eq:F^2}.
From this one may readily obtain 
\[
X=\left[
\begin{array}{c}
0\\1\\1
\end{array}
\right],\
Y=\left[
\begin{array}{c}
0\\1\\0
\end{array}
\right], \
V=\left[
\begin{array}{rr}
1	&0	\\
0	&0	\\
0	&1	
\end{array}
\right],\
W=\left[
\begin{array}{rr}
1	&0	\\
0	&1	\\
0	&-1
\end{array}
\right]
\]
which results in
\[
\tilde{A} =-0.5,\quad 
\tilde{B}_1 = 0,\quad
\tilde{N} = \left[
\begin{array}{rr}
1	&1	\\
-1	&-1	
\end{array}
\right], \quad
\tilde{B}_2 = 
 \left[
\begin{array}{rr}
0.5	\\
0.5	
\end{array}
\right]
\]
and
\[ 
x =\left[\begin{array}{r}
0\\1\\1
\end{array}
\right] z_1 
+ \left[
\begin{array}{rr}
1	&0	\\
0	&0	\\
0	&-1
\end{array}
\right] z_2
\]
}
\end{example}

\newcommand{\half}{\frac{1}{2}}

\begin{remark}
\label{rem:svd}
In general, one can reliably obtain the matrices $X, Y, V, W$ from a {\em singular value decomposition} of $F^k$ where $k$ is greater than or equal to the index of $F$.
Specifically, suppose that
\[
\F ^k =\left[\begin{array}{cc} U_1 & U_2 \end{array} \right]
\left[
\begin{array}{cc}
\Sigma&0\\0 &0
\end{array}
\right]
\left[
\begin{array}{cc}
V_1& V_2
\end{array}
\right]'
\]
is a singular value decomposition of $F^k$ where $\Sigma$ is diagonal  with diagonal elements equal to the nonzero singular values of $F^k$, then
\begin{equation}
X=U_1,
\quad Y= V_1 ,
\qquad V=U_2,
\qquad W=V_2
\qquad
\end{equation}
\end{remark}

\begin{remark}
[Discrete-time systems]
Clearly the results of this paper can be applied to discrete-time descriptor systems described by
the difference algebraic equation
\begin{equation}\label{eq:adescripDis}
Ex(t\!+\!1)=Ax(t)
\end{equation}
where $x(t) \in \mathbb{C}^n$ is the system state at time $t \in \mathbb{N}$ and
 $ E, A \in \mathbb{C}^{n\times n}$. 
 This is because all the results of this paper are only concerned with the pair $(E, A)$ and to obtain discrete-time results just replace $\dot{x}$ with $x(t\!+\!1)$.
 \end{remark}

\section{Conclusions}
In this paper we have obtained order and index reduction results for linear time invariant descriptor systems.
Results are given for both forced and unforced systems as well methods for constructing 
the reduced order systems. Results are also derived that relate our results to existing 
results in the literature. Future work will consider developing similar results for classes of 
nonlinear descriptor systems.


\end{document}